\documentclass[12pt]{iopart}

\usepackage{iopams}  
\usepackage{dsfont}
\usepackage{xcolor}

\newcommand{\ket}[1]{|{#1}\rangle}
\newcommand{\bra}[1]{\langle{#1}|}
\newcommand{\openone}{\mathds{1}}
\newcommand{\bol}[1]{\boldsymbol{#1}}
\newtheorem{thm}{Theorem}
\newcommand{\eg}{{\it e.g.}}

\usepackage{tabulary,graphicx}
\graphicspath{{Images/}}
\usepackage[labelfont=bf]{caption}
\usepackage[labelfont=bf]{subcaption}
\usepackage{wrapfig}

\expandafter\let\csname equation*\endcsname\relax
\expandafter\let\csname endequation*\endcsname\relax

\usepackage{amsmath}

\begin{document}

\title
	{How to enhance quantum generative adversarial learning of noisy information}

	\author{Paolo Braccia$^{1,2,3}$, Filippo Caruso$^{1,3,4}$, and Leonardo Banchi$^{1,2}$}

\address{$^1$ Dipartimento di Fisica e Astronomia, Universit\`a di Firenze, I-50019, Sesto Fiorentino (FI), Italy}
\address{$^2$ INFN, Sezione di Firenze, I-50019, Sesto Fiorentino (FI), Italy}
\address{$^3$ LENS and QSTAR, Via N. Carrara 1, I-50019 Sesto Fiorentino, Italy.}
\address{$^4$ Istituto Nazionale di Ottica CNR-INO, Firenze, Italy.}

%\ead{submissions@iop.org}
\vspace{10pt}
%\begin{indented}
%\item[]August 2020 
%\end{indented}

\begin{abstract}
	Quantum Machine Learning is where nowadays machine learning meets quantum information science. In order to implement this new paradigm for novel quantum technologies, we still need a much deeper understanding of its underlying mechanisms, before proposing new algorithms to feasibly address real problems. In this context, quantum generative adversarial learning is a promising strategy to use quantum devices for quantum estimation or generative machine learning tasks. However, the convergence behaviours of its training process, which is crucial for its practical implementation on quantum processors, have not been investigated in detail yet. Indeed here we show how different training problems may occur during the optimization process, such as the emergence of limit cycles. The latter may remarkably extend the convergence time in the scenario of mixed quantum states playing a crucial role in the already available noisy intermediate scale quantum devices. Then, we propose new strategies to achieve a faster convergence in any operating regime. Our results pave the way for new experimental demonstrations of such hybrid classical-quantum protocols allowing to evaluate the potential advantages over their classical counterparts.
\end{abstract}

%\maketitle

%
%{Quantum algorithms, protocols, and simulations}
%{Quantum computation architectures and implementations}

\section{Introduction}

Machine learning (ML) techniques, besides transforming the way we approach huge-data processing problems, are starting to permeate even non-computer science research and applied sectors, leading to new (big) data-driven strategies, including also several concrete applications in our everyday life as domotic systems, autonomous cars, face/voice recognition, and medical diagnostics.
One of the most outstanding ML results is provided by the generative adversarial networks (GANs) \cite{goodfellow2014generative}, which are models exploiting game-theory theorems \cite{kakutani1941generalization} to learn how to reproduce some given data distribution as close as possible. More specifically, two agents, named as the generator and the discriminator, compete against each other in a zero-sum game, i.e. they play in turns, each turn trying to improve their own strategy in order, respectively, to fool the discriminator and to correctly distinguish real data from generated ones. Under some reasonable assumptions\footnote{The strategy spaces of the agents are compact and convex.} the game has a unique (Nash) equilibrium point, where the generator is able to exactly reproduce the wanted (real) data distribution.

On the other side, in the last few decades quantum information science \cite{watrous2018theory} has focused on how quantum systems store and process information, and how they can be exploited to
implement more efficient protocols than classical ones. This field is currently leading to the first prototypes of quantum devices with some of them already reaching the commercial market, especially in the context of quantum communication and quantum sensing protocols. 
Over the last few years, quantum machine learning (QML) \cite{biamonte2017quantum}, combining ML with quantum information tools, has emerged as one of the most promising applications of near-term quantum devices. Nowadays quantum processors belong to the so-called NISQ (Noisy-Intermediate-Scale-Quantum) era \cite{preskill2018quantum}. Circuits running on such devices are characterized by limited size and depth, and the absence of exact error correction protocol makes them still unsuitable for general-purpose computation. However, they can already be employed for variational algorithms lying at at the core of ML. Indeed, their resilience against noise, together with the assistance of classical algorithms performing the optimization of some variational parameters, leading also to dubbing such algorithms as \textit{hybrid quantum-classical} \cite{zhu2019training,mcclean2016theory,mcclean2017hybrid}, is what makes them feasible on NISQ hardware. Most efforts in QML research are hence devoted to exploit quantum resources as superposition and entanglement to achieve quantum (possibly exponential) speedups in classical ML tasks. In fact, once loaded on a quantum computer, classical data can be processed with linear operations, as those happening in a neural network, which are exponentially faster than what is possible on classical computers \cite{lloyd2010quantum}. Clearly, working in a fully quantum landscape is interesting as well, particularly from the perspective of improving quantum simulations \cite{georgescu2014quantum},   control \cite{dong2010quantum}, metrology \cite{giovannetti2011advances}, and communication \cite{gisin2007quantum} strategies.

In this context, GANs can be successfully generalized to the quantum domain leading to the so-called \textit{Quantum} Generative Adversarial Networks (QGANs) \cite{lloyd2018quantum,dallaire2018quantum}. The aim of QGANs is to learn reproducing the state of a quantum system, usually a register of qubits. A way to achieve this goal, as well as implementing learning algorithms over quantum computers, is to leverage the representation power of parametrized quantum circuits (PQCs). Examples of learning algorithms realized via PQCs are the quantum approximation optimization algorithm  \cite{farhi2014quantum}, and the variational autoencoders \cite{pepper2019experimental} and eigensolvers \cite{peruzzo2014variational}. A detailed review on PQCs and their features, as well as applications, can be found in Ref. \cite{benedetti2019parameterized}. Since PQCs are circuits composed of quantum gates controlled by real tunable parameters, they allow us to steer the output state at our will, by adapting the gate parameters to the measurement outcomes. As for classical neural networks, we can set up a gradient-based strategy to optimally update the parameters.
Moreover, QGANs have been exploited to learn classical distributions of data \cite{romero2019variational,zoufal2019quantum} and to provide a new tool in learning pure states \cite{dallaire2018quantum,benedetti2019adversarial}, whereas mixed states (i.e. noisy information) have been addressed only as ensembles of pure data \cite{hu2019quantum}. However, the latter play a crucial role in the coming NISQ technologies since the environmental noise is unavoidable and usually partially destroys the quantum features as entanglement that do not have a classical counterpart and that are instead mainly responsible for the predicted quantum speedups.

For these reasons we strongly believe that, in order to more deeply understand the performance of QGANs on real hardware, it is remarkably relevant to investigate the scenario of learning mixed quantum states. This is the main focus of this work.
The paper is organized as follows. In Sec. \ref{sec:quant_adv_game} we review the mathematical formalism for QGANs, hence showing why learning mixed states could be an issue. Then, Sec. \ref{sec:training_pqcs} discusses how to implement a QGAN game on PQCs, where we find the emergence of limit-cycle like behaviours (around the equilibrium point) in single-qubit mixed states learning, slowing down the optimization process. Hence, we explore a so-called \textit{optimistic} algorithm that allows to achieve convergence, i.e. destroying the limit cycles. After a discussion about convex optimization over PQCs (Sec. \ref{sec:conv_opt}), conclusions and outlooks are drawn in Sec. \ref{sec:conclusions}.

\section{Quantum adversarial game} \label{sec:quant_adv_game}

\begin{figure}
\centering
\includegraphics[width=0.7\textwidth]{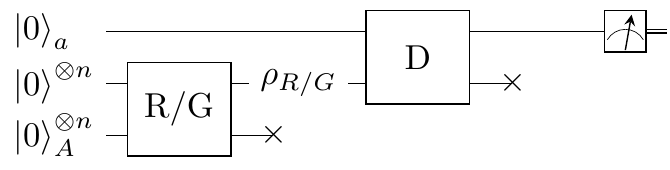}
\caption{QGAN circuit structure for generic $n$-qubit mixed states. The R/G/D blocks are PQCs that are exploited to create real/generated data, and to implement the discrimination process, respectively. The discriminator has access to an ancilla qubit $a$, while the auxiliary qubits $A$ are used by R/G to create mixed states. The {$\times$} symbol over a wire means tracing the degrees of freedom associated to it.}
\label{fig:qgan_struct}
\end{figure}

In GANs the goal of the discriminator (D) is indeed to discriminate {\it real} (R) data from the fake ones generated by the generator (G), while the goal of the latter is to fool the discriminator as much as possible by generating {\it fake} data. Here both real and generated data are modeled as quantum states, respectively described by their density operators $\rho_R$ and $\rho_G$. 
To generate mixed quantum states, one can create a generic pure state (living in a larger Hilbert space) as a quantum circuit by applying quantum gates to a given pure (ground) state, and then tracing out half of the qubit register. The same procedure can be exploited to generate the fake data, but in terms of a PQC where the gate parameters can be tuned. Besides, D applies another PQC to the real or fake state at hand, and entangles it to an additional (ancilla) qubit that later is measured in order to perform the discrimination -- see Fig. \ref{fig:qgan_struct}.

As in the classical case, without any restriction on the operations performed by both agents, 
the game {\it should} end when G is able to perfectly reproduce the real data and,
accordingly, D is unable to correctly discriminate fake data from real ones. Even in the quantum domain, this corresponds to the unique Nash equilibrium of the underlying game~\cite{lloyd2018quantum}. 
Let us point out that, before reaching the equilibrium point, both D and G try to iteratively 
update their strategy to win the game. The D action is modeled via a 
two-outcome positive operator-valued measure (POVM) $\Pi^D_{i}$ whose outcome 
$i\in\{R,G\}$ judges whether the state is real or fake, generated by G. 
Therefore, at each iteration, D has to solve a binary quantum state discrimination task. 
The error in such discrimination process is given by 
the conditional probability of judging real a generated state, i.e. 
$p(R|G)=\Tr[\Pi^D_R \rho_G]$, and by that of judging fake a 
real state, $p(G|R)=\Tr[\Pi^D_G \rho_R]=1-\Tr[\Pi^D_R \rho_R]$. 
Assuming equal {\it a priori} probabilities, and that G's current state is known by D, 
we may define the {\it discrimination error} as $[p(R|G)+p(G|R)]/2$. 
The discriminator strategy during their turn can be formalized as a
minimization of the discrimination error that, without loss of generality, 
can be written as 
\begin{eqnarray}
	{\rm Discriminator}:& ~~~~~~~~  \max_{\Pi_D} \Tr[\Pi_D(\rho_R-\rho_G)], 
	~~~~~~~ {\rm with}~\rho_R,\rho_G~{\rm fixed},
	\label{discriminator}
\end{eqnarray}
where we set $\Pi_D\equiv \Pi^D_R$ to simplify the notation. An analytic 
solution to the above optimization is provided 
by Helstrom's theorem \cite{helstrom1976quantum, holevo1973statistical}, which states that 
the optimum POVM $\{\Pi^D_i\}$ is formed by projectors onto the positive ($\Pi^D_R$) 
and negative ($\Pi^D_G$) subspaces of the operator $\rho_R-\rho_G$.
On the other hand,
the generator's strategy is to fool the discriminator as much as possible 
by reducing their ability to distinguish the real and generated states. 
This results in the following optimal strategy for G
\begin{eqnarray}
	{\rm Generator}:& ~~~~~~~~  \min_{\rho_G} \Tr[\Pi_D(\rho_R-\rho_G)], 
	~~~~~~~ {\rm with}~\rho_R,\Pi_D~{\rm fixed}.
	\label{generator}
\end{eqnarray}
This strategy has a formal analytic solution as  
$\rho_G=\ket{\pi_{\rm max}}\!\bra{\pi_{\rm max}}$,
where $\ket{\pi_{\max}}$ is the eigenvector of $\Pi_D$ with maximum eigenvalue. 
If D is always playing with the optimal Helstrom measurement, then $\rho_G$ is a 
projection onto an eigenstate of $\rho_R-\rho_G$ with positive eigenvalue.

However, it is simple to show that D and G cannot and {\it should not} solve 
the optimization problems \eqref{discriminator} and \eqref{generator} at 
each iteration. Firstly, they {\it cannot} find the optimal solution 
without perfectly knowing, at each iteration, 
$\rho_R$ and the other player's strategy, which contradicts the original scope of the game.
Secondly, they {\it should not} perform such difficult optimization at each round: 
if D and G iteratively play using the solution of 
Eqs.~\eqref{discriminator}-\eqref{generator}, then they never reach the equilibrium 
for mixed states $\rho_R$. 
This is summarized by the following theorem, whose proof is 
straightforward: as discussed above, the solution of \eqref{generator} is always a 
pure state $\rho_G=\ket{\pi_{\rm max}}\!\bra{\pi_{\rm max}}$, and as such $\rho_G\neq \rho_R$ 
in general. 

\begin{thm}
	For mixed states $\rho_R$, the adversarial game fails to converge when D and G iteratively 
	use the strategies \eqref{discriminator}-\eqref{generator}.
\end{thm}

To achieve convergence, each player must {\it slightly} update their strategy 
at each operation \cite{lloyd2018quantum}, 
rather than using Eqs.~\eqref{discriminator}-\eqref{generator}.
Moreover, in the language of Nash equilibria, 
each player is unaware of the adversary's move, and the best they can do is to 
assume that the opponent is playing with the optimal strategy 
and fight against it. 
Setting the {\it score} as the bilinear function
\begin{equation}
	S(\rho_G,\Pi_D)=\Tr[\Pi_D(\rho_R-\rho_G)]~,
	\label{score}
\end{equation}
we see that 
G increases its score whenever D loses the same amount, making this a zero-sum game.
Both the states $\rho_G$ and the measurement operators $\Pi_D$ form a convex set in their respective spaces.\footnote{
Since $\Pi_D$ is part of a POVM it is a positive operator with $\|\Pi_D\|_\infty<1$.}
Therefore, the Nash equilibrium is the result of the minimax problem 
$\min_{\rho_G}\max_{\Pi_D} S(\rho_G,\Pi_D) = \frac12 \min_{\rho_G}  \|\rho_R-\rho_G\|_1 =0$
where the first equality follows from  the Helstrom theorem and the definition of 
the 1-norm \cite{watrous2018theory}. As a result, 
the Nash equilibrium is when the generator is able to perfectly reproduce $\rho_R$,
as originally shown in \cite{lloyd2018quantum}. 
However, how to achieve in practice this equilibrium configuration 
%from ``experience'', i.e. many plays, 
is far from being trivial.

Inspired by the success of gradient-based training of generative adversarial
networks~\cite{goodfellow2016nips}, the most natural approach to play the
quantum adversarial game 
is to use a suitable parametrizations of $\rho_G$ and $\Pi_D$, see e.g. Fig.~\ref{fig:qgan_struct},
and then iteratively 
update these parameters, \eg~via gradient descent 
\cite{lloyd2018quantum}. Using these methods, 
convergence with pure target states $\rho_R=\ket{\psi_R}\bra{\psi_R}$ 
has been obtained in several scenarios 
\cite{dallaire2018quantum,benedetti2019adversarial}, 
while for mixed states convergence was observed with a few extra steps, \eg~by 
setting $\rho_G$ as a random superposition of pure states \cite{hu2019quantum}. 
In spite of these successful examples, gradient-based training may be problematic,
as we numerically investigate in the next section. This is due to the bilinear nature of the score function \eqref{score}. Indeed
%The score ~\eqref{score} is an particular example of bilinear cost functions, which are known to be problematic in the optimization. 
it has been shown that the adversarial optimization of bilinear score functions may display limit cycles when trained with standard gradient descent rules, or even a ``chaotic'' behaviour, see \eg~\cite{mertikopoulos2018cycles,flokas2019poincar} and references therein. 

More precisely, let us consider the simplest case where $\rho_R$ is a single-qubit mixed state. 
The most natural parametrization of $\rho_R$ is via the Bloch vector $\vec r$, namely
$\rho_R = [\openone+\vec r\cdot \vec\sigma]/2$ where $\vec \sigma$ is the vector 
of Pauli matrices and $\vert\vec r \vert\leq 1$. Similarly we parametrize $\rho_G$ with the Bloch vector $\vec g$ and 
$\Pi_D = [d^0\openone+\vec d\cdot \vec\sigma]/2$, where $d^0=\Tr\Pi_D$ and $d^0\geq \vert\vec d \vert$ (because $\Pi_D \geq 0$). With this simple parametrization 
Eq.~\eqref{score} becomes a bilinear form in the Bloch vectors
\begin{equation}
	S(\Pi_D,\rho_G) = \frac{\vec d\cdot(\vec r-\vec g)}2~.
	\label{scoreBLoch}
\end{equation}
The above score function has been extensively investigated in 
Refs.~\cite{daskalakis2017training,zhang2019convergence} 
where the emergence of limit cycles in classical GANs training was shown. 
Nonetheless, Refs.~\cite{daskalakis2017training,zhang2019convergence} focus on 
bilinear problems with linear constraints, while Bloch vectors satisfy a 
non-linear constraint since they live in the Bloch ball. 
This difference may be the reason behind the 
good performance of quantum adversarial learning for pure states
\cite{dallaire2018quantum,benedetti2019adversarial}, 
as pure states lie at the boundary of the Bloch sphere where such non-linear constraints 
are important. However, when dealing with the optimization of highly mixed states, 
which lie well inside the Bloch ball, the presence of the boundary may not affect the optimization, and limit cycles may emerge. 
We summarise this aspect in the following theorem, 
whose proof, adapted from \cite{daskalakis2017training}, can be found in \ref{a:profgd}:

\begin{thm}\label{t:gd}(informal statement):
	Gradient descent/ascent applied to the problem $\min_{\rho_G}\max_{\Pi_D} S(\Pi_D,\rho_G)$ 
	diverges for states far from the surface of the Bloch sphere. 
\end{thm}

\begin{figure}[t!]
  \centering
  \includegraphics[scale=1]{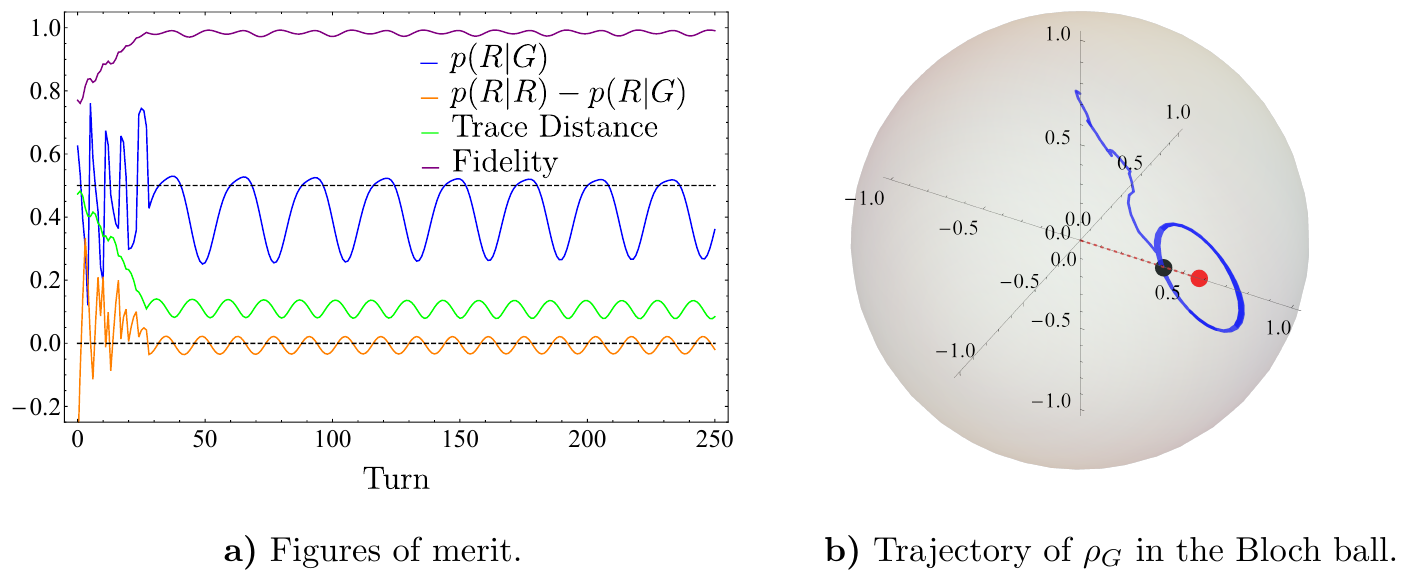}
  \caption{Emergence of a limit cycle when the score \eqref{scoreBLoch} is optimized via gradient descent/ascent, as described in the main text. Here the learning rate of both agents is $\eta=0.1$, and one training turn consists of 5 discriminator's steps followed by a single generator's one.}
  \label{fig:limit_cycle_GD}
\end{figure}

We bring evidence to the previous statement by running a QGAN game in a single
qubit scenario where both D and G are parametrized via their Bloch vectors. We employ 
gradient descent/ascent (GDA) -- namely gradient descent for $\vec g$ and gradient ascent for $\vec d$ --
on the score function \eqref{scoreBLoch} with an added 
penalty term to enforce the constraints on the Bloch vectors, i.e.
$\|\vec g\|\leq 1$ and $\|\vec d\|\leq d^0\leq 2-\|\vec d\|$. 
Results are shown in figure \ref{fig:limit_cycle_GD}, where the limit cycle
behaviour in the trajectory of $\vec g$ is evident.

%%% IMMAGINE

An algorithm dubbed ``optimistic mirror descent'' (OMD) 
has been proposed in Ref.~\cite{daskalakis2017training} to escape 
from the limit cycles that emerge in the minimax optimization of bilinear cost
functions \eqref{scoreBLoch}. In the 
next section we show that, although perfect limit cycles may not exist for more 
complex parametrizations of $\rho_G$ and  $\Pi_D$, a simple gradient descent/ascent update 
may produce a ``chaotic'' behaviour, where convergence is not observed. We find instead that 
convergence is obtained via OMD. 

\section{Training with parametric quantum circuits}\label{sec:training_pqcs}

Motivated by the capabilities of current
noisy intermediate-scale quantum hardwares \cite{preskill2018quantum}, 
common PQCs are based on single qubit gates
controlled by tunable real parameters, \eg~qubit rotations around a fixed axis
with variable angle, and non-parametric two-qubit gates,
typically CNOTs.\footnote{The Controlled-NOT (CNOT) operation applies Pauli's $X$ to the target qubit if the control one is found in $\vert 1 \rangle$ and does nothing otherwise.} Sequences of single and two-qubit gates  are then stacked in a layered 
fashion. 
When sketched down, it is easy to see a certain resemblance
with classical neural networks, with the parameters playing the role of
the weights and biases of the latter. Indeed, it turns out that, just as a neural network can represent any function given the proper depth and structure, an accurately built PQC can approximate any unitary mapping over the input quantum register. Indeed, CNOT gates and single-qubit rotations are universal 
for quantum computation, i.e. they can be composed to simulate any unitary evolution to the desired accuracy \cite{nielsen2010quantum}. 
PQCs  can be designed to comply with nowadays NISQ hardware, by 
adapting the two-qubit (entangling) gates to the connectivity of the
experimental realization of the quantum processor, and by limiting the
circuit's depth to fight decoherence. 
Since learning tasks ultimately boil down to
the problem of minimizing a certain loss/score function of the model
parameters, we can employ PQCs as quantum learning models and tune them through
a feedback loop with a classical optimizer. This is the standard framework of
hybrid quantum-classical variational approaches \cite{mcclean2016theory}, and we
will use this scheme in our analysis.

Here, we will ultimately be concerned with the problem
of learning a mixed state via a QGAN game. Since every mixed state can be written as
a pure state in a larger Hilbert space (Fig.~\ref{fig:qgan_struct}), 
we build the generator via the following PQC with classical parameters $\bol\theta_G$
\begin{eqnarray}
	\rho_G &= \Tr_A \left[\ket{\psi_{GA}(\bol\theta_G)}
	\bra{\psi_{GA}(\bol\theta_G)}\right]~, 
		~~~ \ket{\psi_{GA}(\bol\theta_G)} &= 
	U(\bol{\theta}_G)\ket{0}^{\otimes 2 n}~,
	 \label{Gpara}
 \end{eqnarray}
where both $A$ and $\rho_G$ contain $n$ qubits, and $U(\bol\theta_G)$ is the unitary 
operator corresponding to the PQC. 
Similarly, since every measurement operator can be written as a projective
measurement onto a larger Hilbert space (Fig.~\ref{fig:qgan_struct}), 
we define the discriminator's POVM  with classical parameters  $\bol\theta_D$ as
\begin{equation}
	\Pi_D = \Tr_{a} \left[U(\bol\theta_D)^\dagger(\openone_D\otimes \ket0
{}_{a}\bra0)U(\bol\theta_D)\right]~.
	\label{Dpara}
\end{equation}
where $a$ is a single auxiliary qubit. 
This measurement can be interpreted as follows: 
first apply a PQC $U(\bol\theta_D)$ entangling the system with an 
auxiliary qubit $a$, then measure the qubit $a$ in the computational basis. If the outcome $0$
is detected, then we guess that the state is the real state, otherwise (outcome 1) the state is judged as fake.

\subsection{Circuits Ansatz}
\begin{figure}[t!]
\centering
\includegraphics[scale=1]{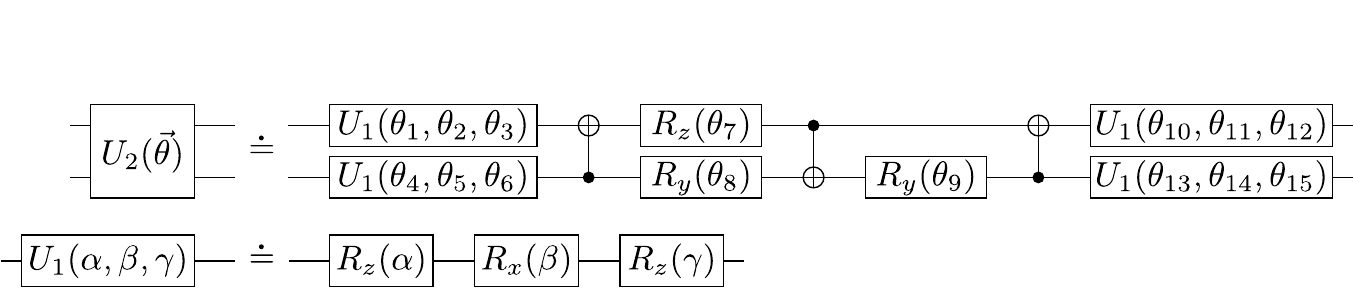}
\caption{The building block of G and D circuits with 3 CNOTs and 15 single-qubit rotations. $U_1$ and $U_2$ implement elements of ${\rm SU}(2)$ and  ${\rm SU}(4)$, respectively, where $R_i$ are rotations around the $i$-th axis.}
\label{circ_block}
\end{figure}

Following Refs.~\cite{benedetti2019adversarial,shende2004minimal}, G and D circuits are built by repeating a two-qubit block which implements a
generic unitary $U\in{\rm SU}(4)$. As shown in Fig.~\ref{circ_block}, this building block is
composed of 15 single-qubit rotations and 3 CNOT gates.
One block allows to generate every two-qubit pure state. For larger registers, we apply this block to each pair of consecutive qubits, thus obtaining a layer. Layers are then concatenated in a staggered pattern.

Throughout this manuscript, the gradients needed to train the QGAN are analytically computed 
through the \textit{parameter shift rule} 
\cite{mitarai2018quantum,schuld2019evaluating,banchi2020measuring}. Since the parametrized part of our circuit consists of single-qubit rotations, the gradient of a function $f(\vec{\theta})=\langle O(\vec{\theta})\rangle$ is obtained as
\begin{equation}\label{parm_shift_rule}
\frac{\partial f}{\partial \theta_i} = \frac{1}{2}\left[f\left(\vec{\theta} +\frac{\pi}{2}\vec{e}_i \right) - f\left(\vec{\theta} - \frac{\pi}{2}\vec{e}_i\right) \right] \; ,
\end{equation}
where $\vec{e}_i$ is the unit vector in the $i$-th direction.

\begin{figure}[t!]
\centering
\includegraphics[scale=1]{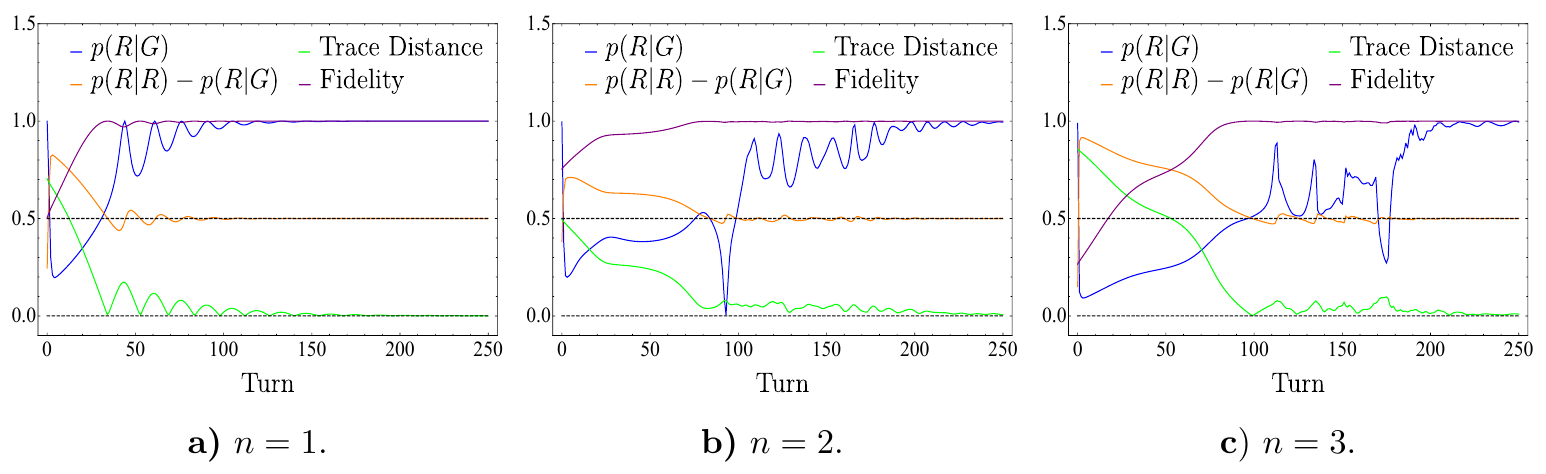}
\caption{Convergence of QGANs for learning pure states of $n$ qubits. Both agents rely on standard gradient descent/ascent optimization. The target and the initial fake states are randomly chosen on the Bloch sphere.}
\label{fig:pure_states_triple}
\end{figure}

We first focus on learning pure states, for which it is known that QGANs converge. Indeed, Fig. \ref{fig:pure_states_triple} does further confirm it in terms of the relevant figures of merit, as the
score function value, the probability $p(R\vert G)$ of D labelling fake data as
true, the trace distance $d=\frac{1}{2}\|\rho_G - \rho_R\|_1$ 
between the generated state and the target one, and their
fidelity $F=\left[\Tr\sqrt{\sqrt{\rho_G}\rho_R\sqrt{\rho_G}}\right]^2$. In our simulations the target {\it real} data are 
random pure states of $n$ qubits, with $n=1,2,3$, 
obtained via a PQC with the same structure of the one used for G, but with random fixed parameters. Here, training is carried out via alternately updating D
and G via a single gradient descent/ascent step. We have tried different optimizers, always observing a qualitatevely similar convergence behaviour%
\footnote{Hyperparameters such as the learning rates might be further fine-tuned in order
to improve convergence speed, but this is beyond the aim of this manuscript.}.

\subsection{Emergence of limit cycles}

\begin{figure}[t!]
    \centering
	\includegraphics[scale=1]{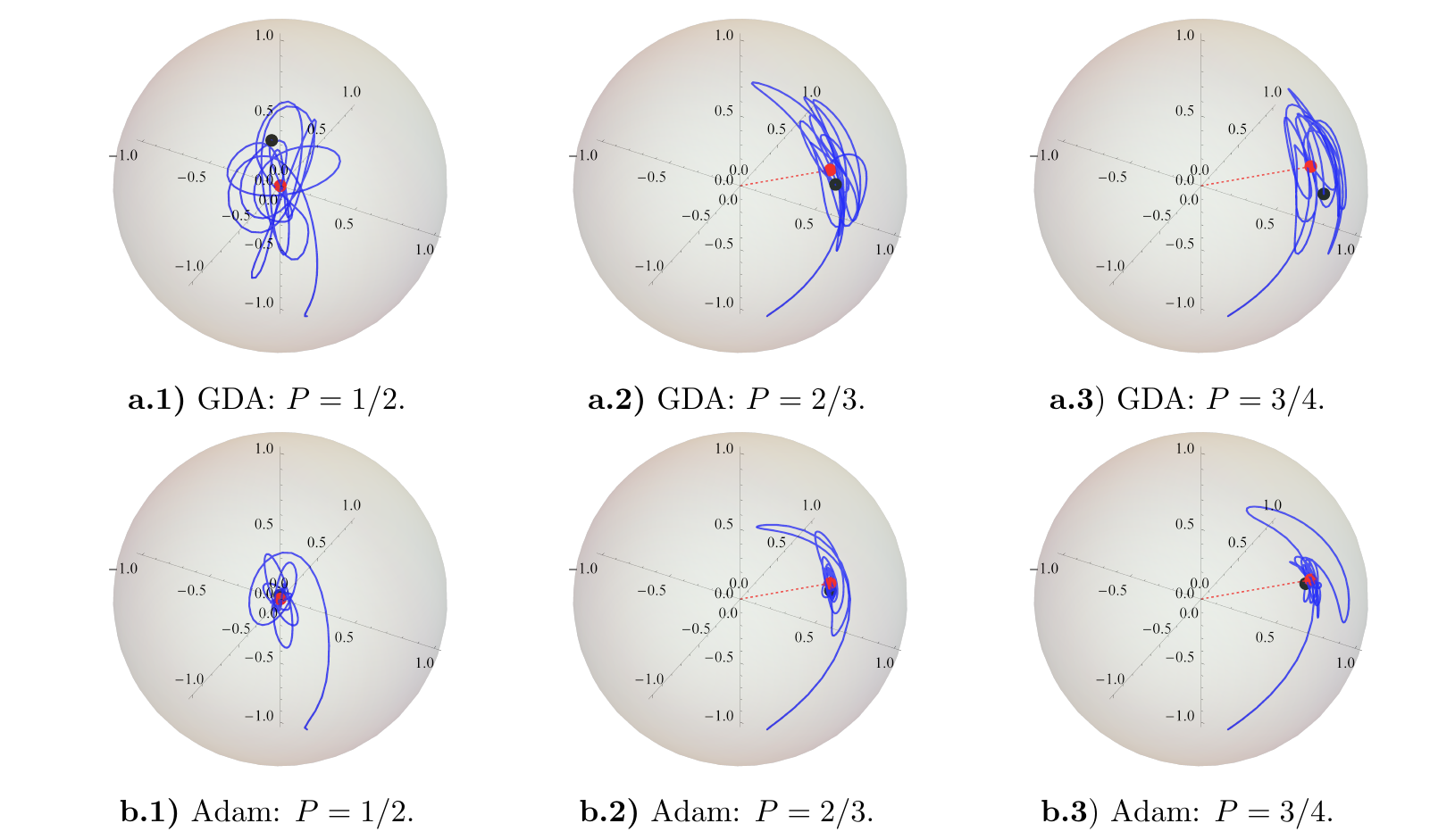}
	\caption{Limit-cycle-like behaviour of QGANs when learning mixed states for different values of the purity $P$. As optimizers, we use GDA (top row) and Adam (bottom). None of them display convergence, although the latter has a less pronounced oscillating behaviour. In all these cases the initial configurations of G and D correspond to the same random parameters. These trajectories have been obtained by running the QGAN for 250 total turns, where each turn comprises 10 optimization steps for D followed by 1 for G. Lastly, the learning rate of GDA is set to 0.1, whereas that of Adam is 0.05.}
\label{fig:limit_cycles_triple}
\end{figure}

\begin{figure}[h!]
  \centering
  \includegraphics[scale=1.1]{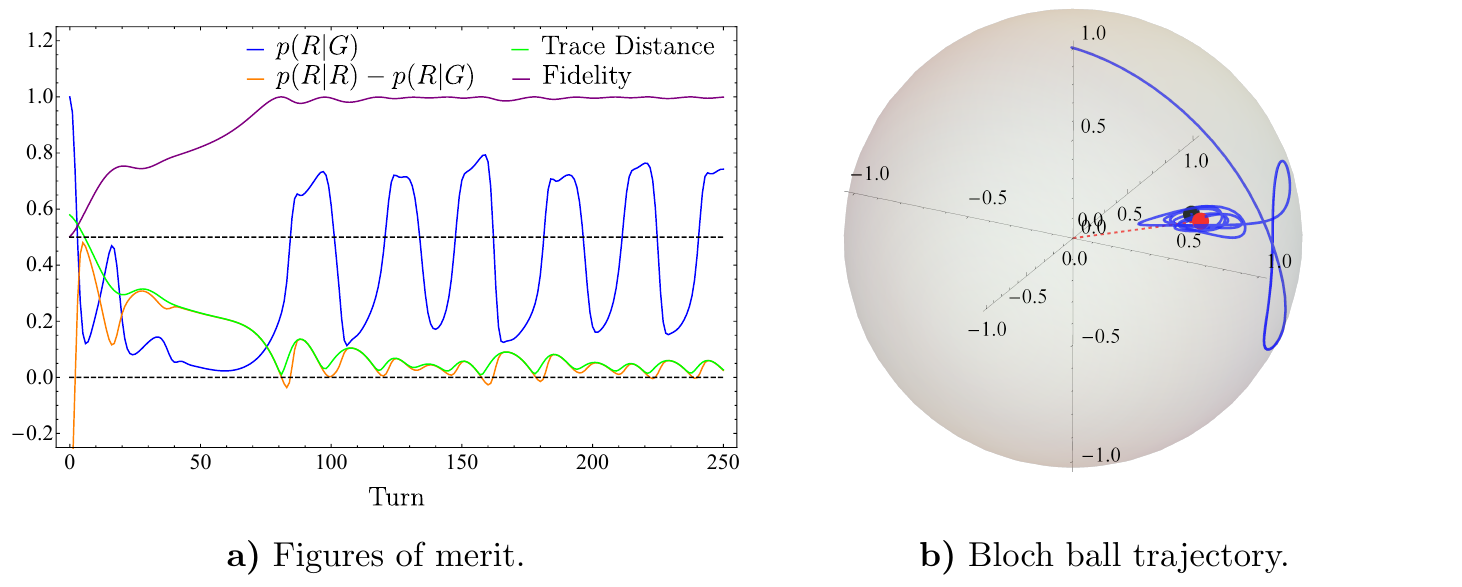}
  \caption{Training behaviour for learning a mixed state with $P=3/4$ under the same conditions of \ref{fig:limit_cycles_triple} with Adam and the tailored agents, but a different initial configuration.}
  \label{fig:lazo_Adam_tailored}
\end{figure}

We now generalize the above analysis to the more interesting case of learning mixed states. They have been so far addressed as an ensemble of
orthogonal pure states \cite{hu2019quantum}, while here they are created by tracing out half of the qubits register. 
Our results are summarized in Fig. \ref{fig:limit_cycles_triple}, where we show the learning process for mixed states of the form $\rho_R=\frac{I}{2}+\frac{a}{2\sqrt{2}}(\sigma_x+\sigma_y)$ with purity $P=\tr{\rho^2}=\frac{1+a^2}{2}$, ranging from the
completely mixed one $P=1/2$ to $P=3/4$. The selected optimizers are the previously defined GDA and \textit{Adam}, i.e. one of the best performing optimization algorithm for ML \cite{kingma2014adam}.
As we can see in Fig. \ref{fig:limit_cycles_triple}, none of the chosen optimizers allows to reach convergence, even by changing the values of the optimization hyperparameters. However, comparing these results with the ones in Fig.
\ref{fig:limit_cycle_GD}, we can see that for PQCs the limit-cycle behaviour disappears because the score function is no longer bilinear.
\begin{figure}[t!]
\centering
\includegraphics[scale=1]{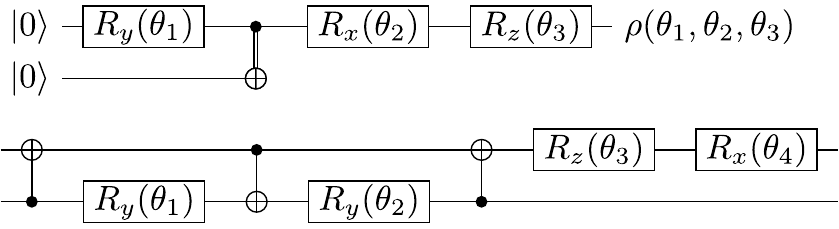}
\caption{The minimal circuits for G (top), and D (bottom).}
\label{fig:minimal_circs}
\end{figure}
Let us point out that in Fig. \ref{fig:limit_cycles_triple} we have an overparametrization because $D$ and $G$ use 15 parameters each, whereas a general single-qubit POVM has 4 real degrees of freedom only, and a single qubit mixed state has 3. For this reason we devise two tailored circuits in order to achieve
a minimal parametrization for both D and G (see
figure \ref{fig:minimal_circs}), as in the following:
\begin{equation}
\rho_G(\bol \theta)=\frac{1}{2}
\begin{pmatrix}
1 + c (\theta_1) c (\theta_2) & c (\theta_1) s (\theta_2) \left(s (\theta_3)+i c (\theta_3)\right) \\
c (\theta_1) s (\theta_2) \left(s (\theta_3) - i c (\theta_3)\right) & 1 - c (\theta_1) c (\theta_2)
\end{pmatrix},
\end{equation}
and
\begin{equation}
\Pi_D(\bol\theta)=\frac{1}{2}
\begin{pmatrix}
1+c\left(\theta_1 + \theta_2\right)c\left(\theta_4\right) & s(\theta_4)\left(c(\theta_1)s(\theta_3) - i c(\theta_2)c(\theta_3)\right) \\
s(\theta_4)\left(c(\theta_1)s(\theta_3) + i c(\theta_2)c(\theta_3)\right) & 1 - c\left(\theta_1 - \theta_2\right)c\left(\theta_4\right)
\end{pmatrix} \; ,
\end{equation}
with $\cos(\theta) \equiv c(\theta)$ and $\sin(\theta) \equiv s(\theta)$.
Even with these tailored circuits, convergence is not achieved as numerically reported in Fig. \ref{fig:lazo_Adam_tailored}.
Moreover, by using simple gradient descent method, we still observe limit cycles (not shown in figure).

% limit cycle Adam TAILORED

\subsection{Training with optimism}

\begin{figure}[t!]
\centering
\includegraphics[width=\linewidth]{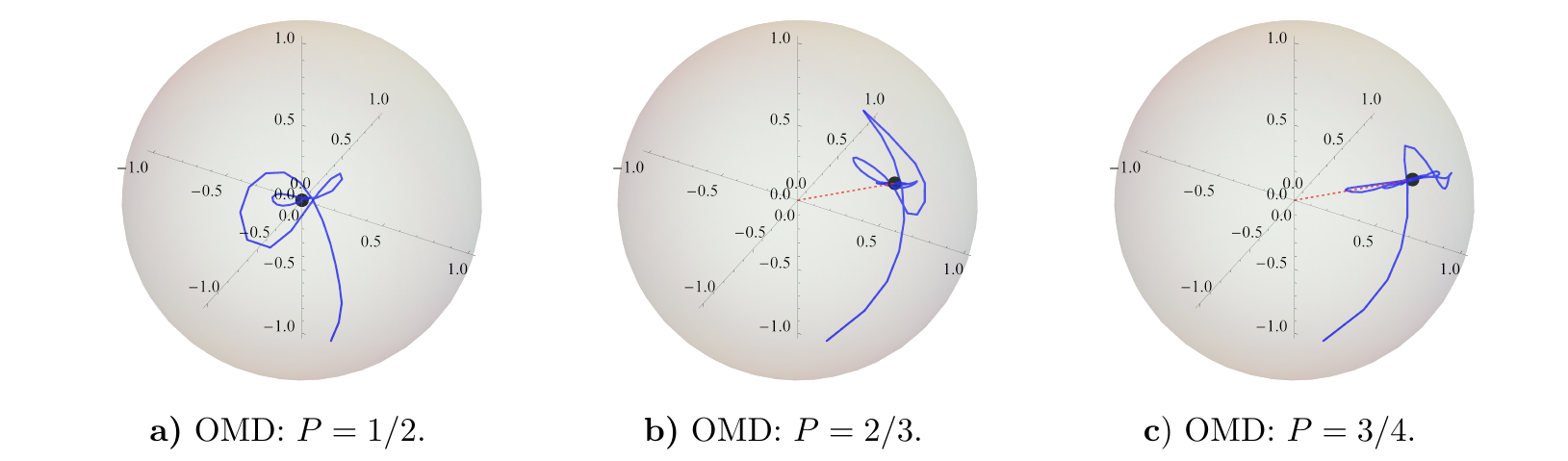}
\caption{Convergence of QGANs for learning mixed states via OMD under the same conditions used in Fig. \ref{fig:limit_cycles_triple}.}
\label{fig:convergence_omd_triple}
\end{figure}

In standard GANs competing players are usually unaware of the opponent's strategy. However, each player may try to guess the opponent's move in order to improve its strategy. This is the building concept of the \textit{Optimistic Mirror Descent} (OMD) optimization algorithm \cite{rakhlin2013online}, which was shown to fix convergence issues, namely
limit cycles, of classical GANs with bilinear score functions -- see Ref. \cite{daskalakis2017training}. However, there it has been used to enhance convergence also in the case of non-bilinear score functions. Motivated by these results, we now show that OMD works successfully also for our QGANs -- see Fig. ~\ref{fig:convergence_omd_triple}. More specifically, the OMD-based update rule for the score function of Eq.~\eqref{score}, 
$S(\bol\theta_D,\bol\theta_G) := S(\Pi_D(\bol\theta_D),\rho_G(\bol\theta_G))$, 
reads 
%for a set of parameters $\{\vec \theta \}$ and for a loss function $L$ to be minimized, reads
\begin{align}
	\bol \theta^{t+1}_D &= \bol \theta^{t}_D + 2\eta_D \nabla_{\bol \theta_D}S(\bol\theta^D_{t},\bol\theta^G_{t})
	- \eta_D \nabla_{\bol \theta_D}S(\bol\theta^D_{t-1},\bol\theta^G_{t-1}),
\\
\bol \theta^{t+1}_G &= \bol \theta^{t}_G - 2\eta_G \nabla_{\bol \theta_G}S(\bol\theta^D_{t+1},\bol\theta^G_{t})
	+ \eta_G \nabla_{\bol \theta_G}S(\bol\theta^D_{t},\bol\theta^G_{t-1}),
\end{align}
where $\eta_{D/G}$ are the learning rates for D and G. Notice that this rule corresponds to the scenario where D is optimized first.

%%% FIGURA TRIPLA GD E ADAM AL VARIARE DELLA PURITY
%
\section{Convex optimization}\label{sec:conv_opt}
Since PQCs are not the only way of modelling quantum states, here we present a non-parametric method, hereafter dubbed ConvexQGAN,
to solve the minimax problem 
$\min_{\rho_G}\max_{\Pi_D} S(\rho_G,\Pi_D)$ 
using the formalism of convex optimization presented in Ref. \cite{banchi2020convex}. 
Since $\{\rho_G\}$, $\{\Pi_D\} $ are both convex sets and 
$S(\rho_G,\Pi_D)$ is bilinear, when we iteratively fix either 
$\rho_G$ or $\Pi_D$ we always obtain a convex function over a convex set.
Therefore, by adapting the Frank-Wolfe algorithm from Ref.~\cite{banchi2020convex},
we may write the following update rules at the $k$th step
\begin{eqnarray}
	\Pi_D^{k+1} &= (1-\alpha_k) \Pi_D^{k} + \alpha_k \ket{D_k}\bra{D_k}~,
	\label{convexD}
	\\
	\rho_G^{k+1} &= (1-\beta_k) \rho_G^{k} + \beta_k \ket{G_k}\bra{G_k}~,
	\label{convexG}
\end{eqnarray}
where $\alpha_k$ and $\beta_k$ are decaying learning rates, e.g. typically 
$\alpha_k=\beta_k=\frac{2}{k+2}$, the state
$\ket{G_k}$ is the eigenvector with smallest eigenvalue of 
$\nabla_{\rho_G} S(\rho_G^{k},\Pi_D^{k}) = - \Pi_D^{k}$, while 
$\ket{D_k}$ is the eigenvector with smallest eigenvalue of 
$-\nabla_{\Pi_D} S(\rho_G^{k},\Pi_D^{k}) = -(\rho_R-\rho_G^{k})$.
Although the update rules directly follow from the Frank-Wolfe algorithm, 
we highlight here an interesting result from the physics points of view. 
The states $\ket{D_k}$ are 
elements of Helstrom measurement to optimally distinguish the 
real state $\rho_R$ from the current fake state $\rho_G^k$. 
As such, it is tempting to consider a different strategy with $\alpha_k=1$ at 
each iteration step. The downside of the latter approach is that the measurement operator highly fluctuates 
between different steps. However, for $\alpha_k=1$ we get $\ket{D_k}=\ket{G_k}$ so 
Eq.~\eqref{convexG} gets a clear operational meaning. 
The generator's state is iteratively updated 
with one of the states entering in the Helstrom optimal measurement. 
This reminds us the original optimization from Eq.~\eqref{generator}, but without its convergence 
issues for mixed states. Indeed, the update rule of Eq.~\eqref{convexG} allows the generation of 
mixed states, unlike in Eq.~\eqref{generator}.

Finally we propose a physics inspired alternative by observing that, for small $\beta_k$,
Eq.~\eqref{convexG} can be interpreted as an imaginary time evolution 
\begin{equation}
\rho_G^{k+1} \propto e^{\beta_k H_k} \rho_G^{k} e^{\beta_k H_k}~, ~~~~~
H_k=\ket{G_k}\bra{G_k}~,
	\label{imagG}
\end{equation}
where after the imaginary evolution we need to normalize the state such as $\Tr[\rho_G^{k+1}]=1$.  
\begin{figure}[t!]
	\centering
	\includegraphics[width=0.7\linewidth]{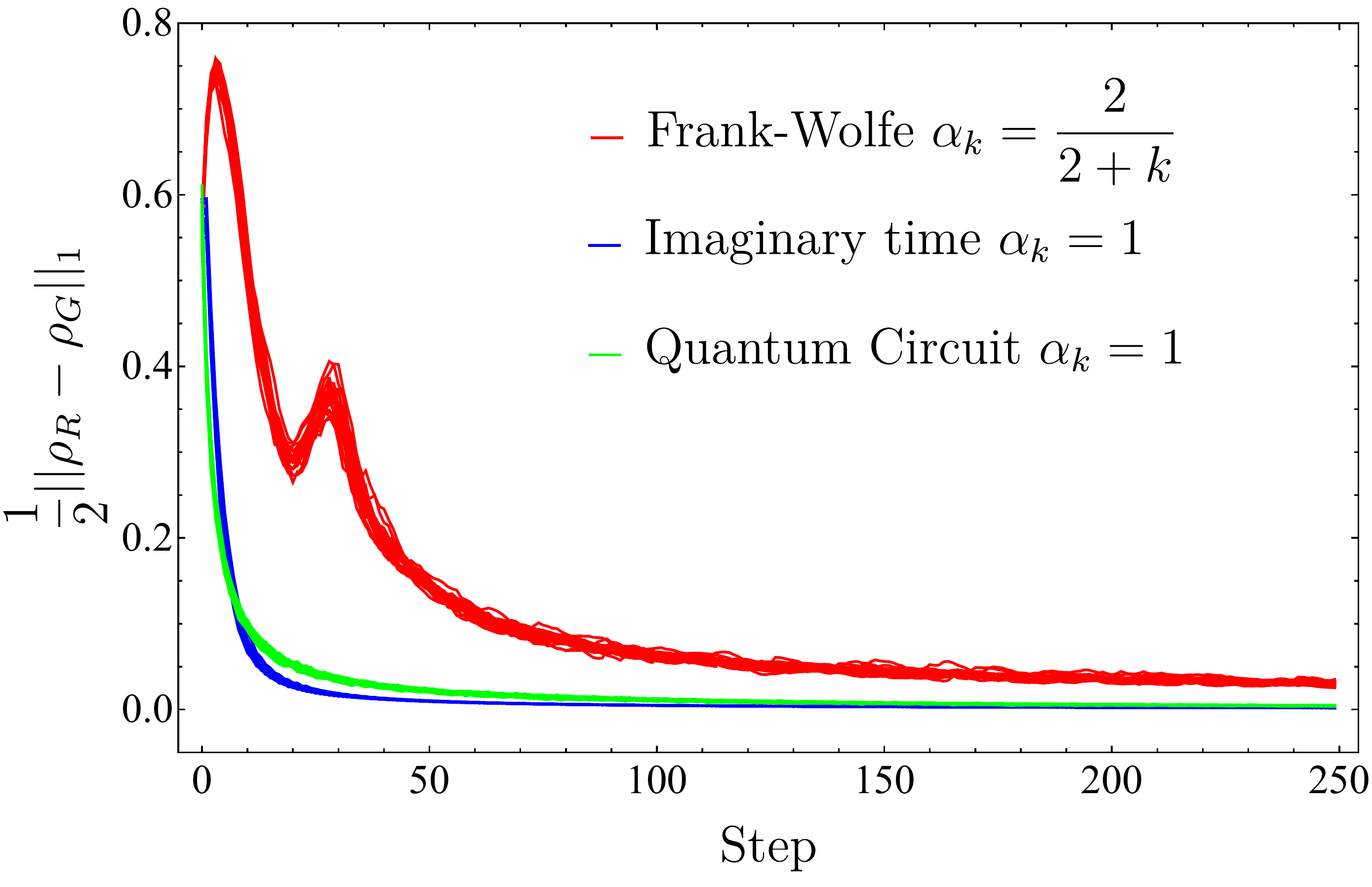}
	\caption{ConvexQGAN: Learning different random mixed states using either the Frank-Wolfe iteration 
		\eqref{convexD}-\eqref{convexG}, the imaginary time evolution \eqref{imagG} or the quantum 
		circuit update from \eqref{circuitG}. The cases with 
		$\alpha_k=1$ in \eqref{convexG} correspond to Helstrom measurements. 
		For each algorithm, 20 lines are plotted for different random initial states. All simulations 
		are for 5-qubit states. 
	}
	\label{fig:convexgan}
\end{figure}
The gradient-based Frank-Wolfe  algorithm	\eqref{convexD}-\eqref{convexG} and 
the imaginary time iteration \eqref{imagG} are numerically studied in Fig.~\ref{fig:convexgan} 
for random 5-qubit states with full-rank. For the imaginary time iteration we use $\alpha_k=1$, so $\ket{G_k}=\ket{D_k}$
in \eqref{imagG}. We observe in Fig.~\ref{fig:convexgan} that the imaginary time evolution, together 
with the optimal Helstrom measurement at each step, significantly outperforms the Frank-Wolfe iteration, 
both in terms of speed and accuracy. 

ConvexQGAN methods show fast convergence towards the equilibrium configuration, but they require 
eigendecompositions of the state at each step. This operation is simple for classical computers 
as long as the Hilbert space is sufficiently small. To extend this operation to larger systems, 
we now discuss how to write an update like in Eq. \eqref{imagG}, but using a quantum circuit. For this purpose, 
we use the following quantum map, which is at the heart of the quantum density matrix exponentiation 
algorithm \cite{lloyd2014quantum},
\begin{equation}
	\mathcal E^{\pm t}_\sigma[\rho] = \Tr_2[e^{\pm i t\, {\rm SWAP}} \rho\otimes\sigma e^{\mp it \, {\rm SWAP}}]
	= \cos(t)^2 \rho + \sin(t)^2 \sigma \pm \frac i2 \sin(2t)[\rho,\sigma]~,
\end{equation}
where SWAP is the swap operator. Applying this map twice with different signs, one has 
\begin{equation}
	\mathcal E^{-t}_\sigma \circ \mathcal E^{+t}_\sigma[\rho] = \cos(t)^4\rho+\sin(t)^2(1+\cos^2t)\sigma + \frac14\sin(2t)^2 
	[[\rho,\sigma],\sigma]~.
\end{equation}
Therefore, setting $I^t_\sigma[\rho] = 	\mathcal E^{-t}_\sigma \circ \mathcal E^{+t}_\sigma[\rho]$ 
and $t_k$ such that $\cos^4t_k=1-\beta_k$, namely $\beta_k\approx 2t_k^2$,
we get 
\begin{equation}
	\rho_G^{k+1} =	
	I^{t_k}_{H_k}[\rho_G^k] = (1-\beta_k)\rho_G^k + \beta_k ( H_k + H_k\rho_G^k+\rho_G^k H_k - 2 H_k\rho H_k)+\mathcal O(\beta_k^2)~,
	\label{circuitG}
\end{equation}
where $H_k$ was defined in \eqref{imagG}. The latter update rule is akin to
a mixture of \eqref{convexG} and \eqref{imagG}, but it has the advantage that it 
can be explicitly evaluated 
as a quantum circuit applied to $\rho_k$ and two copies of the state $\ket{G_k}$. 
As shown in Fig.~\ref{fig:convexgan}, the performance obtained with the update rule \eqref{circuitG} 
is similar to that of imaginary time evolution. Therefore, 
if the states $\ket{G_k}$ can be efficiently prepared, for instance via strategies like 
the Helstrom classifier circuit from \cite{lloyd2020quantum}, then the above update rule can be used for QGAN training in a quantum computer. 

\section{Conclusions}\label{sec:conclusions}

In this work we have studied  the convergence of quantum generative adversarial learning for 
mixed states. We have first showed that ``learning'' via simple gradient descent/ascent updates,
or even via more advanced methods such as the Adam optimizer, may be problematic when 
the target state is mixed. 
We have attributed such convergence issues to the bilinear nature 
of QGAN's score function.  Indeed, it is known from classical
GAN literature that the optimization of such score functions leads to limit cycles, 
where the generator gets stuck into, cycling around the target solution 
without ever reaching it. 
We have observed that states obtained via PQCs, 
such as those commonly implemented in nowadays available NISQ devices, are less affected by 
the emergence of limit cycles, but may nonetheless display a ``chaotic'' behaviour during training, without achieving convergence. 

We have then proposed new algorithms for reliable training of QGANs, which always achieve
convergence in our numerical simulations. The first algorithm, suitable for PQCs, is based on the adaptation of optimistic mirror descent, i.e. a gradient-based technique allowing provable convergence 
with bilinear score functions. 
The second algorithm is based on convex optimization techniques, and is especially suited for 
non-parametric states that are iteratively updated via a suitably designed, yet non-parametric, quantum circuit. 

Thanks to our theoretical and numerical analysis, we believe that the proposed algorithms should 
work better than previously used techniques for QGAN training, especially when highly mixed states are involved. 
Having good training heuristics for learning mixed states will help in
leveraging their higher representation power, as well as in providing us with a way to study noisy quantum maps. 
Indeed, a next necessary step for the classification of QGANs capabilities is the analysis of their performance against noise. This path has been paved in
\cite{anand2020experimental}, where it was shown that adversarial schemes share
the noise robustness of other known hybrid quantum-classical variational
algorithms \cite{kandala2017hardware,dong2019robust,barkoutsos2020improving,gentini2019noise}.
Lastly, from the physics point of view, since QGANs perform an implicit state tomography, we believe that, by
further endowing our scheme with the ability to process entangled copies of the target state, performance will be enhanced. It is an open question whether an adversarial strategy may take over some current metrology scheme
\cite{giovannetti2011advances}, by providing faster
and more efficient strategies for sensing and system certification. 
Our results shed new light on how hybrid classical-quantum QML protocols might be exploited in already available experimental platforms with potential promising applications in quantum computing and noise sensing.

\ack{
	L.B. acknowledges support by the program ``Rita Levi Montalcini'' 
	for young researchers, Grant No. PGR15V3JYH,  funded by 
	``Ministero dell'Istruzione, dell'Universit\`a e della Ricerca (MIUR)''. F.C. was financially supported by the Fondazione CR Firenze through the projects QUANTUM-AI, the PATHOS EU H2020 FET-OPEN Grant No. 828946, and the Florence University Grant Q- CODYCES.
}

\appendix
\section{Proof of Theorem~\ref{t:gd}}\label{a:profgd}

The following proof is adapted from \cite{daskalakis2017training}. 
We use the Bloch parametrization \eqref{scoreBLoch}, which we rewrite for simplicity 
as $S(\vec d,\vec g) = \vec d \cdot (\vec r-\vec g)$ ignoring the constant factors.
Gradient descent/ascent applied to $\min_{\vec g}\max_{\vec d} S(\vec d,\vec g)$ results in the 
update rule 
\begin{eqnarray}
	\label{gradlinear1}	
	\vec d_{t+1} &= \vec d_t+\eta (\vec r-\vec g_t),  \\
	\vec g_{t+1} &= \vec g_t-\eta (-\vec d_t),
	\label{gradlinear2}	
\end{eqnarray}
where $\eta$ is a suitably small ``learning rate'' and $t$ is the iteration step. 
The unique fixed 
point is with $\vec g=\vec r$ and $\vec d=0$, which physically corresponds to 
perfect generation $\rho_G=\rho_R$ and impossibility to distinguish real 
from generated data $\Pi_D\propto\openone$. We evaluate the distance from this 
fixed point as $\Delta_t = \|d_t\|_2^2 + \|r-g_t\|_2^2$, where $\|\cdot\|_2$ is 
the $\ell_2$-norm. From \eqref{gradlinear1}-\eqref{gradlinear2} we get 
\begin{eqnarray}
\|\vec d_{t+1}\|_2^2 &= \|\vec d_{t+1}\|_2^2  +2 \eta \; \vec d_t\cdot(\vec r-\vec g_t) + \eta^2 \|\vec r-\vec g_t\|_2^2,
\\
	\|\vec r -\vec g_{t+1}\|_2^2 & = \|\vec r -\vec g_{t}\|_2^2 
-2 \eta \; \vec d_t\cdot(\vec r-\vec g_t) + \eta^2 \|\vec d_{t+1}\|_2^2,
\end{eqnarray}
and accordingly 
\begin{equation}
	\Delta_{t+1} = (1+\eta^2) \Delta_t~,
\end{equation}
namely the distance from the equilibrium point increases at each iteration. 
We point out that the above proof is valid only when we neglect the physical constraints on the Bloch vectors. The latter are however important for pure states or for states near the surface of the Bloch sphere.

\section*{References}
\bibliographystyle{iopart-num}
\bibliography{quganbib}

\end{document}